\documentstyle[aps]{revtex}      %prints out galley style

\begin{document}

\draft

\title{Comment on ``Strangeness enhancement in $p+A$ and \\
       S$+A$ interactions at energies near 200 $A$ GeV"}

\author{Marek Ga\'zdzicki}

\address{Institut f\"ur Kernphysik, Universit\"at Frankfurt, \\
         August-Euler-Str. 6, D-60486 Frankfurt, Germany}
 
\author{Ulrich Heinz}

\address{Institut f\"ur Theoretische Physik, Universit\"at Regensburg, \\
         D-93040 Regensburg, Germany}

\date{\today}

\maketitle

\begin{abstract}

 We argue that the recent analysis of strangeness production in 
 nuclear collisions at 200 $A$ GeV/$c$ performed by Topor Pop {\it et 
 al.} \cite{To:95} is flawed. The conclusions are based on an 
 erroneous interpretation of the data and the numerical model 
 results. The term ``strangeness enhancement" is used in a 
 misleading way.

\end{abstract} 


\narrowtext

In a recent publication Topor Pop and collaborators (``TP*") 
\cite{To:95} discuss the production of strange particles in nuclear 
collisions at CERN SPS energies within microscopic models. Their 
analysis and conclusions are mainly based on the comparison of the 
data from the NA35 experiment \cite{Al:94} with the HIJING model 
\cite{Wa:94}.  

In this Comment we wish to point out that the analysis procedure used 
by TP* is problematic, both with respect to the interpretation of the 
data and to the way these are compared to the model. The results 
presented in TP* do not support the conclusions drawn by the authors.

We wish to first comment on problems with the procedure of TP* 
and afterwards on inconsistencies in the interpretation of their 
results.

{\bf I.} 
In the abstract of Ref.~\cite{To:95} one reads: ``The HIJING model is 
used to perform a {\it linear} extrapolation from $pp$ to $AA$". In 
order to justify this procedure a comparison between $p{+}p$ data 
\cite{Ga:91} and the HIJING model is done in Section III A of 
Ref.~\cite{To:95}. From this comparison TP* conclude: ``We note 
that the {\it integrated} multiplicities for neutral strange 
particles $\langle \Lambda \rangle$, $\langle \overline{\Lambda} 
\rangle$ and $\langle K^0_S \rangle$ are reproduced at the level of 
three standard deviations for $pp$ interactions at 200 GeV. However, 
the values for $\langle \overline{p} \rangle$ and $\langle 
\overline{\Lambda} \rangle$ are significantly over predicted by the 
model". This, together with Figs. 1a, 2a of Ref.~\cite{To:95}, is 
taken as evidence that the HIJING model is sufficiently accurate in 
its reproduction of $\Lambda$ and kaon production in $p{+}p$ 
collisions to allow for a meaningful extrapolation to $p+A$ and $A+A$ 
data.  

An inspection of Table I in Ref.~\cite{To:95} leads, however, to the 
opposite conclusion: The yields of $\langle \Lambda \rangle$ and 
$\langle K^0_S \rangle$ are significantly overpredicted by  HIJING 
(6--7 standard deviations \cite{fn1} for $\Lambda$ and 9--10 standard 
deviations for $K^0_S$). The yields of $\overline{p}$ and 
$\overline{\Lambda}$ seem also to be overpredicted by HIJING but 
they agree with the model within 3 standard deviations. 

Furthermore, Fig.~1a of Ref.~\cite{To:95} shows that HIJING also 
fails to reproduce the $\Lambda$ rapidity spectrum in $p{+}p$ 
collisions. In fact it was shown previously by one of the authors of 
TP* that the HIJING model underpredicts the stopping of baryons in 
nuclear collisions \cite{Gy:95}. This biases the form of the 
$\Lambda$ rapidity distribution producing characteristic 
forward--backward peaks (see Figs. 1a--d in Ref.~\cite{To:95})
which are not observed or (in the case of $p{+}p$ collisions) are 
significantly lower in the data.  

We therefore conclude that the HIJING model has severe shortcomings 
in its reproduction of the $p{+}p$ data which eliminate it as a 
candidate for ``a {\it linear} extrapolation from $pp$ to $AA$". A 
detailed discussion of the effect of  strangeness enhancement in $AA$ 
cannot be reliably based on the comparison with this model.  

{\bf II.}
In Sec.~III C of Ref.~\cite{To:95} the rapidity and the transverse mass 
(momentum) spectra of strange particles are compared with the HIJING 
and VENUS models. This analysis is misleading and the ensuing 
discussion of the transverse mass spectra is meaningless
since the authors do not restrict the calculated spectra to 
the rapidity acceptance of the experimental data. The failure to 
properly account for experimental acceptances is also reflected in
the following misleading statement in the introduction of 
Ref.~\cite{To:95}: ``In addition, there have been substantial changes 
in the final published data \cite{Al:94} relative to earlier 
comparisons to preliminary data \cite{Ba:90,St:91}". In fact the 
experimental data published in \cite{Ba:90,St:91} are  
consistent with the recently published results \cite{Al:94}; the 
differences are  due to an enlarged acceptance and statistics of strange 
particles in S+S collisions \cite{Xi} following modifications of the NA35 
set--up \cite{Al:94} which then also allowed for an analysis of 
strangeness production in S+Ag and S+Au collisions \cite{Al:94}.  

We expect that taking into account the experimental acceptances will 
result in lowering the model points by a factor of up to about 3. One 
should also note that different data sets have significantly 
different acceptances. These acceptance cuts can not be neglected in 
the analysis of the transverse mass spectra, as done by TP*, since 
they influence both the spectral shapes and the local yields of 
strange particles.  

Let us now comment on the conclusions drawn by TP* from their 
results. They write \cite{To:95}: ``Our main conclusion therefore is 
that strangeness enhancement is a nonequilibrium effect clearly 
revealed in the lightest ion interactions". This conclusion is based 
on the following TP* observation: ``The strangeness enhancement in 
the minimum bias $p$+S is striking because the number of target 
nucleons struck by the incident proton is on the average only 2."  
The conclusion and its justification do NOT follow from the TP* 
analysis. Its origin can be traced as follows: TP* compare the
$\Lambda$ rapidity distribution for $p$+S interactions with the HIJING 
model (Fig. 1b in Ref.~\cite{To:95}). They observe that HIJING 
underpredicts the $\Lambda$ yield at midrapidity. This disagreement 
is called by TP* the  observation of a strangeness 
enhancement in $p$+S interactions. However, as argued above, the 
underprediction of midrapidity $\Lambda$'s by HIJING is a direct 
consequence of its weak baryon stopping. Fig.~1b of Ref.~\cite{To:95} 
shows very clearly that the {\em under}prediction of midrapidity 
$\Lambda$'s in $p$+S collisions by HIJING is accompanied by an {\em 
over}prediction of $\Lambda$'s in the proton fragmentation region.  
This reflects the incorrect description of baryon stopping in $p{+}A$ 
collisions by the model. As such it has nothing to do with an 
enhanced production of strange particles.  

Since the discovery of anomalously high production of strangeness in 
central nucleus--nucleus collisions at AGS and SPS energies 
\cite{Ha:89,Ga:89} ``strangeness enhancement" has been defined in a 
model independent way as an increase of the ratio between the {\em 
total} multiplicity of strange quarks (particles) and that of 
non-strange quarks (particles) when going from nucleon--nucleon 
interactions to nuclear  collisions. It can be 
quantified \cite{Wr:85,Bi:92} by studying the change of the 
strangeness suppression factor $\lambda_S$ which is commonly used in 
the elementary particle physics, or by analysing the factor $E_S$ 
introduced in \cite{Al:94} in order to avoid experimental problems in 
the evaluation of $\lambda_S$ for nuclear collisions. The compiled 
data on strangeness production in $p$+$p$ \cite{Ga:91} and $p$+$A$ 
\cite{Bi:92,Al:94} interactions lead to the conclusion that {\em 
there is no strangeness enhancement in $p$+$A$ interactions at 200 
GeV/c}. This statement is based on 8 independent measurements of 
strange and non-strange particle production in $p$+$A$ interactions, 
with $A$ ranging from Mg to Au. The NA35 data from $p$+S collisions 
alone lead also to the same conclusion.  

On the other hand, {\em strangeness enhancement is observed in 
central nucleus--nucleus collisions} at all studied collision 
energies \cite{Ga:95a} (e.g. for central S+S and S+Ag collisions at 
200 $A$ GeV/c). Due to the model deficiencies discussed in this 
Comment, the analysis of TP* does not allow to trace the mechanism 
for this enhancement. The statement of TP* that it is ``clearly a 
nonequilibrium effect" which prohibits the use of ``simplistic 
fireball models" for its interpretation has not been proven. None of 
the known microscopic and kinetic models based on hadronic and string 
dynamics (including HIJING and VENUS) is able to reproduce the 
strangeness enhancement effect observed at CERN SPS energies 
\cite{Ga:95}.  

\begin{center}
{\bf ACKNOWLEDGMENTS}
\end{center}
We would like to thank  P. Seyboth and H. Str\"obele for
discussions.


\begin{thebibliography}{[9]}
\bibitem{fn1}
  The experimental error of $\langle \Lambda + \Sigma^0 \rangle$ 
  quoted in Table I in \cite{To:95} should be 0.01 instead of 
  0.015, see Tables 5 and 7 in Ref. \cite{Ga:91}.
\end{thebibliography}
\end{document}